\documentclass[twocolumn,10pt,prl, aps,superscriptaddress,showpacs,reprint,amsmath,amssymb]{revtex4-1}
\usepackage{subfigure}

\usepackage{dcolumn}
\usepackage{bm}
\usepackage{color}
\usepackage{latexsym}
\usepackage{psfrag,graphicx}
\usepackage{color}
\usepackage[english]{babel}
\usepackage{epstopdf}
\usepackage{appendix}
\usepackage{changes}
\usepackage[colorlinks=true,citecolor=myblue,linkcolor=myred]{hyperref}
\usepackage{lastpage}
\def\be{\begin{equation}}
\def\ee{\end{equation}}
\def\ba{\begin{align}}
\def\enda{\end{align}}
\def\bi{\begin{itemize}}
	\def\ei{\end{itemize}}

\def\beq{\begin{equation}}
\def\eeq{\end{equation}}
\def \bea{\begin{eqnarray}}
\def \eea{\end{eqnarray}}

 \def\ee{\mathord{\rm e}}

\usepackage{subfigure}

\usepackage{dcolumn}
\usepackage{bm}
\usepackage{color}
\usepackage{psfrag,graphicx}
\usepackage{color}
\usepackage[english]{babel}
\usepackage{epsf} 
\usepackage{amsfonts}
\usepackage{epstopdf}
\usepackage{appendix}
\usepackage{changes}

\usepackage{ifpdf}
\usepackage{bm}
\usepackage{color}
\usepackage{pstricks}
\usepackage{figsize}
\usepackage{rotating}
\usepackage{subfigure}
\usepackage{rotating}
\usepackage{ifthen}
\usepackage{ifpdf}
\usepackage{gensymb}
\usepackage{amssymb}
\usepackage{bm}
\usepackage{moresize}
\ifpdf

\usepackage{graphicx}
\usepackage{epstopdf}
\else
\usepackage{graphicx}
\usepackage{epsfig}
\fi

\definecolor{mygrey}{gray}{0.35}
\definecolor{myblue}{rgb}{0.2,0.2,0.8}
\definecolor{myzard}{cmyk}{0,0,0.05,0}
\definecolor{mywhite}{rgb}{1,1,1}
\definecolor{myred}{rgb}{1,0.,0.3}
\definecolor{MATblue}{rgb}{0 0.4470 0.7410}
\definecolor{MATorange}{rgb}{0.8477 0.3242 0.0977}
\definecolor{MATgreen}{rgb}{0 .75 0}

\renewcommand{\ee}{{\rm e}}

\begin{document}
	
	\title[Short Title]{Strong angular-momentum mixing in ultracold atom-ion excitation-exchange}
	
	\author{Ruti Ben-Shlomi} \address{Department of Physics of Complex Systems, Weizmann Institute of Science, Rehovot 7610001, Israel}
	\author{Romain Vexiau}\address{Laboratoire Aim$\acute{e}$ Cotton, CNRS/Universit$\acute{e}$ Paris-Sud/ENS Paris-Saclay/Universit$\acute{e}$ Paris-Saclay, Orsay Cedex, France}
	\author{Ziv Meir}\altaffiliation{Current address: Department of Chemistry, University of Basel, Klingelbergstrasse 80, CH-4056 Basel, Switzerland} \address{Department of Physics of Complex Systems, Weizmann Institute of Science, Rehovot 7610001, Israel}
	\author{Tomas Sikorsky}\altaffiliation{Current address: Atominstitut - E141, Technische Universität Wien, Stadionallee 2, 1020 Vienna, Austria} \address{Department of Physics of Complex Systems, Weizmann Institute of Science, Rehovot 7610001, Israel}
	\author{Nitzan Akerman} \address{Department of Physics of Complex Systems, Weizmann Institute of Science, Rehovot 7610001, Israel}
	\author{Meirav Pinkas} \address{Department of Physics of Complex Systems, Weizmann Institute of Science, Rehovot 7610001, Israel}
	\author{Olivier Dulieu}\address{Laboratoire Aim$\acute{e}$ Cotton, CNRS/Universit$\acute{e}$ Paris-Sud/ENS Paris-Saclay/Universit$\acute{e}$ Paris-Saclay, Orsay Cedex, France}
	\author{Roee Ozeri}\address{Department of Physics of Complex Systems, Weizmann Institute of Science, Rehovot 7610001, Israel}
	
	\date{July 10th, 2019}
	\begin{abstract}
		{Atom-ion interactions occur through the electric dipole which is induced by the ion on the neutral atom. In a Langevin collision, in which the atom and ion overcome the centrifugal barrier and reach a short internuclear distance, their internal electronic states deform due to their interaction and can eventually alter. Here we explore the outcome products and the energy released from a single Langevin collision between a single cold  $^{88}$Sr$^{+}$ ion initialized in the metastable $4d^2D_{5/2,3/2}$ states, and a cold  $^{87}$Rb atom in the $5s^2S_{1/2}$ ground state. We found that the long-lived $D_{5/2}$ and $D_{3/2}$ states quench after roughly three Langevin collisions, transforming the excitation energy into kinetic energy. We identify two types of collisional quenching. One is an Electronic Excitation-Exchange process, during which the ion relaxes to the $S$ state and the atom is excited to the $P$ state, followed by energy release of $\sim$ 3000 K$\cdot$k$_B$. The second is Spin-Orbit Change where the ion relaxes from the higher fine-structure $D_{5/2}$ level to the lower $D_{3/2}$ level releasing $\sim$ 400 K$\cdot$k$_B$ into kinetic motion. These processes are theoretically understood to occur through Landau-Zener avoided crossings between the different molecular potential curves. We also found that these relaxation rates are insensitive to the mutual spin orientation of the ion and atoms. This is explained by the strong inertial Coriolis coupling present in ultracold atom-ion collisions due to the high partial wave involved, which strongly mixes different angular momentum states. This inertial coupling explains the loss of the total electronic angular-momentum which is transferred to the external rotation of nuclei. Our results provide deeper understanding of ultracold atom-ion inelastic collisions and offer additional quantum control tools for the cold chemistry field.} 
	\end{abstract}                                          
	\maketitle
	
	One of the most challenging contemporary goals in the research of cold atoms is to study a single ultracold collision and gain full quantum control over its outcomes and reaction pathways. Ultracold atom-ion systems offer excellent tools to investigate such cold inelastic collisions. The high fidelity with which trapped ions and atoms can be coherently controlled enables their preparation in well-defined internal and motional states prior to the collision. Similarly, the state of ions and atoms following the collision can be analyzed with great precision. Several collisional processes have been recently studied in ultracold atom-ion mixtures. Examples include charge-exchange reactions \cite{SchmidPRL2010,HallPRL2011,Tacconi2011,Haze2015,RatschbacherNat2012,RellergertPRL2011,GrierPRL2009}, spin-exchange and spin-relaxation \cite{Ratschbacher2013,phase2018,Tscherbul2016, Tomas2018}, molecular formation \cite{HallPRL2011,Hall2013,Hall2013_2,dasilva2015} as well as three-body recombination \cite{Harter2012}.
	
	Most of the experiments mentioned above were carried out with both the atom and the ion prepared in their electronic ground state. However, if one of the colliding partners is prepared in an excited electronic state, more inelastic channels become available, leading to richer dynamics, but also to various challenges in the theoretical interpretation. Inelastic atom-ion collisions with the ion prepared in an optically excited metastable state were recently studied \cite{HallPRL2011,RatschbacherNat2012,Saito2017PRA,Staanum2004}. These metastable states of ions are important since they serve as excited states in optical frequency standards \cite{Margolis2004} and quantum computing \cite{Schmidt2003} applications. In previous experiments, the lifetime of these excited states and the outcome products of such collisions were measured. However, the lack of ability to measure the energy of the outgoing collision products prevented detailed experimental identification of the underlying mechanisms.
	
	Here we report on the observation of the quick relaxation of the long-lived $4d\,^2D_{3/2,5/2}$ (hereafter referred to as the $D_{3/2}$ and $D_{5/2}$ states, for simplicity),  atomic levels of $^{88}$Sr$^+$ when colliding with ultracold $^{87}$Rb atoms in the ground state. We found that the $^{88}$Sr$^+$ ion quickly quenches from the $D_{5/2}$ and $D_{3/2}$ states, after three Langevin collisions on average. By measuring the distribution of the kinetic energies of the ion after the quench, we identify two relaxation processes: Electronic Excitation-Exchange (EEE) and Spin-Orbit Change (SOC). Using accurate molecular structure calculations we show that these processes are induced by short-range interactions, manifested by avoided crossings of the corresponding Potentials Energy Curves (PECs) of the RbSr$^+$ molecule. The SOC channel can be turned on or off by initializing the ion in the $D_{5/2}$ or $D_{3/2}$ level, respectively. We further observe that these processes are insensitive to the initial mutual spin orientation of the atom and ion, indicating a high degree of angular-momentum mixing during the collision. With our typical atom-ion center-of-mass kinetic energy of 0.5 mK$\cdot$k$_B$, up to $\sim $20  partial waves participate in the collision. This high rotational motion generates strong Coriolis forces which are responsible for this angular-momentum mixing. 
	\begin{figure}
		\centering
		\includegraphics[width=0.5\textwidth]{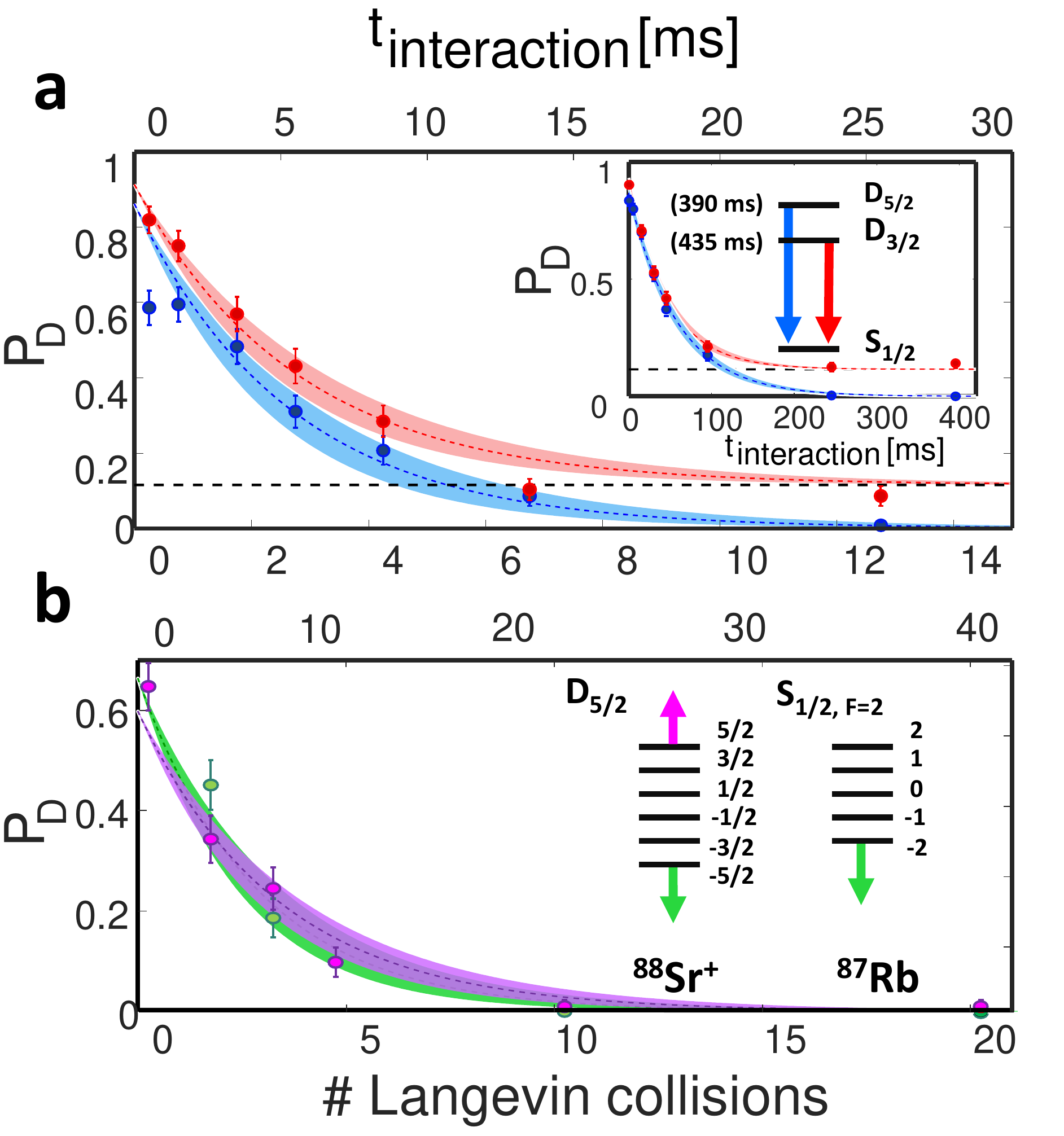}
		\caption[D state lifetime in presence of atoms]{\footnotesize{\textbf{Decay of the ion population from the metastable $D_{3/2,5/2}$ states.} \textbf{(a)} Population decay from the $D_{5/2}$ (blue) and  $D_{3/2}$ (red) states in the absence (inset) and in the presence of atoms. Error bars correspond to one $\sigma$ standard deviation. The natural lifetimes of these levels are 390 and 435 ms correspondingly. The fitted dashed curves are the solutions to rate equations yielding exponential decay. The shaded areas represent exponential-fit confidence bounds of one $\sigma$ on the decay rate. Population is seen to decay after $\simeq$ 3 Langevin collisions on average
				\textbf{(b)} Population decay from the $D_{5/2}$ level, when $^{87}$Rb atoms are polarized in the $m_{F}=-2$ hyperfine Zeeman state, and the ion is polarized to the $m_{J_{e}}=-5/2$ state (green curve), or the $m_{J_{e}}=+5/2$ state (purple curve). These two configurations correspond to the atom and ion internal angular-momentum being parallel vs. anti-parallel.}
			\label{quench}}
	\end{figure}
	
	In our experiment, 20,000 ultracold (3 $\mu$K) $^{87}$Rb atoms are trapped in an optical dipole trap (ODT) and overlapped with a single cold (40 $\mu$K) $^{88}$Sr$^{+}$ ion which is trapped in a linear rf Paul trap \cite{apparatus2018}. Imperfect micro-motion compensation and the heating of the ion during collisions with the atoms \cite{Meir2016} set the typical center-of-mass energy, during the first few collisions, to be roughly 0.5 mK$\cdot$k$_B$. We initially prepared the ion in one of the internal, metastable, excited $D_{5/2}$ or $D_{3/2}$ levels (with natural lifetimes of 390~ms and 435~ms, respectively) and the atoms in the $S_{1/2}$ ground state. We overlapped the cloud of atoms with the single trapped ion for different interaction times, allowing one to few, Langevin collisions on average, with a typical Langevin collision rate of 0.5 kHz (see Supplementary Material for details). After the interaction, the atoms were released from the trap and were measured by absorption imaging after a short time-of-flight. Finally, the ion state and temperature were interrogated using a 422 nm laser close to resonance with its $5s\,^2S_{1/2}-4p\,^2P_{1/2}$ dipole-allowed transition (or $S_{1/2}- P_{1/2}$ in short, see Supplementary Material). All experiments were repeated with an interlaced comparison without the presence of atoms.
	
	In a first experiment, we measured the probability that the ion remains in the metastable $D$ level as a function of the interaction time. Figure~\ref{quench}a shows this probability in the case where the ion was initialized in the $m_{J_{e}}=-5/2$ Zeeman sub-level of the $D_{5/2}$ manifold using electron-shelving with a narrow line-width laser (blue data points), and in the case where the ion was initialized in the $D_{3/2}$ level with unpolarized spin using optical-pumping on the strong $S_{1/2}-P_{1/2}$ dipole-allowed transition (red data points). In both cases the atoms were prepared in the $F=1$ hyperfine manifold of their $S_{1/2}$ ground state without any preferred spin-polarization. When the ion was initialized in the $D_{3/2}$ level the probability did not asymptotically approach zero due to an artifact in our $D_{3/2}$ level population measurement (see Supplementary Material for details). The collisional quenching rate $\Gamma_{Q}$ is extracted from a fit to the solution of a rate equation, shown by the dashed lines (see Supplementary Material). Here, the rate at which both $D$ levels decay in the absence of atoms, due to off-resonance scattering of photons from the ODT laser beam at $1064$ nm, had to be taken into account. This rate was calibrated in a separate measurement shown in the inset of Fig.~\ref{quench}a. The extracted rates at which the two levels decayed from the $D_{5/2}$ and $D_{3/2}$ states are $\Gamma_{Q}=2.6(4)$ and $\Gamma_{Q}= 2.8(3)$ Langevin collisions on average,  respectively.
	
	Events in which the ion decayed from the $D$ level were identified either by state-selective fluorescence, indicating that the ion is cold and in the $S_{1/2}$ ground state, or by observing that the ion has heated up significantly and no fluorescence is recorded even after optically-pumping the ion to the $S_{1/2}$ ground state due to very large Doppler shifts (see Supplementary Material). By comparing the measured collisional quench rate to the rate at which hot ion events were recorded, we concluded that the ion considerably heated up every time it collisionally quenched from the $D$ level, indicating that the quench is non-radiative and releases the internal electronic energy into atomic motion. 
	\begin{figure*}[ht]
		\centering
		\includegraphics[width=0.9\textwidth]{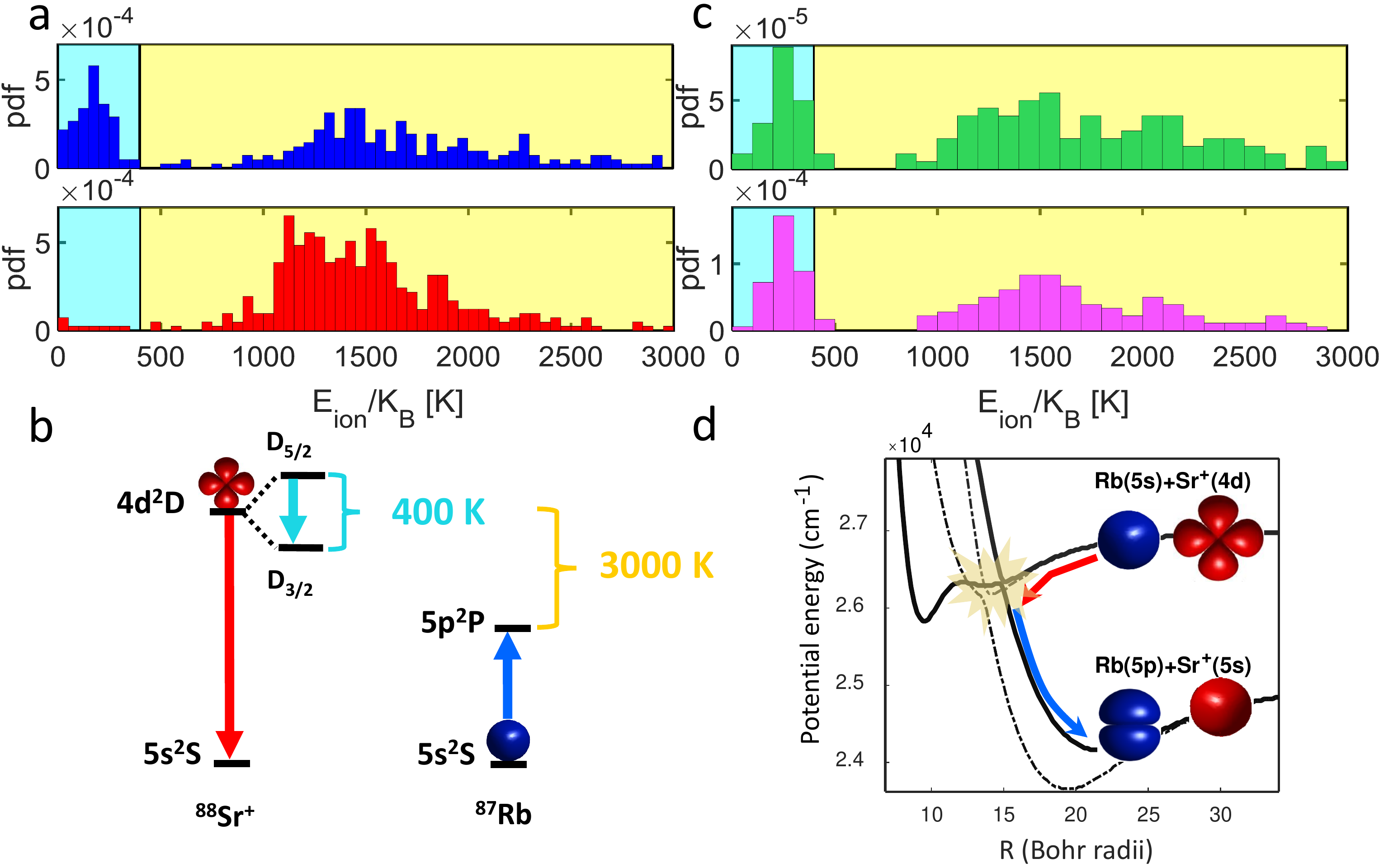}
		\caption{\textbf{Histograms of the energy released in quench events.} \textbf{(a)} Quenching events from the $D_{5/2}$ (top, blue) and $D_{3/2}$ states (bottom, red). Both exhibit a broad energy distribution around 1500 K$\cdot$k$_B$ marked by the yellow background, corresponding to the EEE channel. The $D_{5/2}$ histogram displays a second peak around 200 K$\cdot$k$_B$, marked by the cyan background, corresponding to the SOC channel. 
			The collision probability for this experiment is low, therefore a few events at temperatures lower than 500 K for the $D_{3/2}$ case, are likely the result of sympathetic cooling. \textbf{(b)} Relevant energy levels of $^{88}$Sr$^{+}$ and $^{87}$Rb. The $4d\,^2D$ level in Sr$^+$ splits into two spin-orbit components separated by $\sim$ 400K K$\cdot$k$_B$, which are coupled by the SOC channel (cyan arrow). The energy difference of the EEE channel (red and blue arrows) is $\sim$ 3000 K$\cdot$k$_B$. \textbf{(c)}  Quenching rates from the $D_{5/2}$ state, when the internal angular-momentum states of $^{88}$Sr$^{+}$ and $^{87}$Rb are prepared in an aligned (top, green, referred to as $\downarrow_{i} \downarrow_{a}$), and anti-aligned (bottom, purple, referred to as $\uparrow_{i} \downarrow_{a}$) configuration (see Fig.~\ref{quench}b). The ratio between the corresponding rates are:  $\Gamma^{\downarrow_{i}\downarrow_{a}}_{SO}/\Gamma^{\downarrow_{i}\downarrow_{a}}_{EEE}=31(4) \%$, and  $\Gamma^{\uparrow_{i}\downarrow_{a}}_{SO}/\Gamma^{\uparrow_{i}\downarrow_{a}}_{EEE}=47(7)   \%$. \textbf{(d)} Schematic illustration of the quenching process leading to EEE, and of the relevant (RbSr)$^{+}$ molecular PECs. The process is induced by a localized avoided crossing between two curves (solid - $^1\Sigma^{+}$, dashed - $^3\Sigma^{+}$, see also Fig.~\ref{adiabatic2}b), leading to a large kinetic energy release of $\sim$ 3000 K$\cdot$k$_B$.}
		\label{Hist6}.
	\end{figure*}
	
	Surprisingly, we did not observe any signal suggesting enhanced charge-exchange reaction rate when the ion was initialized in the excited $D$ state. The measured upper bound on the probability of undergoing a charge-exchange reaction during a Langevin collision when the ion is in this excited state is $P \ll 1\times10^{-4}$. This is in contrast to all previous observations in which charge exchange was observed to be dominating for ions excited to metastable levels 
	\cite{RatschbacherNat2012,Saito2017PRA,HallPRL2011,RellergertPRL2011,GrierPRL2009,SchmidPRL2010}.
	
	We next turned to investigate the dependence of the quench rate on the mutual spin orientation of the ion and atom. We interlaced the experiment between initializing the ion in the $D_{5/2},\ m_{J_{e}}=-5/2$ and the $D_{5/2},\ m_{J_{e}}=+5/2$ spin states. The atoms were prepared in the stretched spin state  $F=2, m_{\textrm{F}}=-2$. Here, when the ion and atom spins are aligned, the total electronic angular-momentum of the atom-ion complex is $3\hbar$. Since there is no lower electronic energy level that carries that much angular-momentum, non-radiative decay would imply the transfer of angular-momentum from the electronic degrees of freedom to rotation of nuclei. Such coupling between internal and rotation angular-momenta in ultracold collisions is usually weak \cite{Mies1996}, so that we would expect the parallel-spin quench rate to be suppressed as compared with the anti-parallel spin case. Figure~\ref{quench}b shows the two measured decay curves. Using a fit to the measured data we found that the ion quenches from the $D_{5/2}$ level after $\Gamma_{Q}=2.4(2)$ (${m_{J_{e}}}=-5/2$) and $\Gamma_{Q}=3.0(3)$ (${m_{J_{e}}}=5/2$) Langevin collisions on average. Strikingly, the quenching rate does not depend on the mutual electronic angular-momentum orientation of the atom and ion, implying strong electronic angular-momentum transfer to molecular rotation.
	
	In order to investigate the spectrum of kinetic energies released in the quench we performed Single-Shot Doppler Cooling Thermometry (SSDCT) on the ion (see Supplementary Material and Ref.\cite{single-shot, Direct-sympathetic}). SSDCT measures the initial energy of the ion prior to Doppler cooling, independently for every repetition of the experiment, provided that this energy is above $\simeq$ 10 K$\cdot$k$_B$. Here we set the interaction time to be sufficiently short ($\sim$ 0.5 ms), to avoid multiple elastic collisions leading to sympathetic cooling of the ion after the quench event.
	
	The upper blue histogram in Fig.~\ref{Hist6}a, shows the distribution of energies measured following a quench of the $D_{5/2}$ level. Two clear separate energy distributions emerge. One peak around $\simeq$ 200  K$\cdot$k$_B$ and another around $\simeq$ 1500 K$\cdot$k$_B$. These energies can easily be associated with decay channels by considering the energy differences between atomic levels shown in Fig.~\ref{Hist6}b. Decay from the entrance channel Rb($S_{1/2}$)+Sr$^+$($D_{5/2}$) to the Rb($P_{1/2,3/2}$)+Sr$^+$($S_{1/2}$) channels (EEE channels),  releases a kinetic energy of $\simeq$ 3230 K$\cdot$k$_B$ and $\simeq$ 2890  K$\cdot$k$_B$ respectively. This energy is almost equally divided between $^{87}$Rb and $^{88}$Sr$^+$ owing to their nearly equal masses, resulting in the peak around $\simeq$ 1500 K$\cdot$k$_B$. Similarly, a quench to the Rb($S_{1/2}$)+Sr$^+$($D_{3/2})$ channel (SOC channel), releases of $\simeq$  402 K$\cdot$k$_B$, resulting in the peak around $\simeq$ 200 K$\cdot$k$_B$.
	\begin{figure}[h]
		\includegraphics[width=0.5\textwidth]{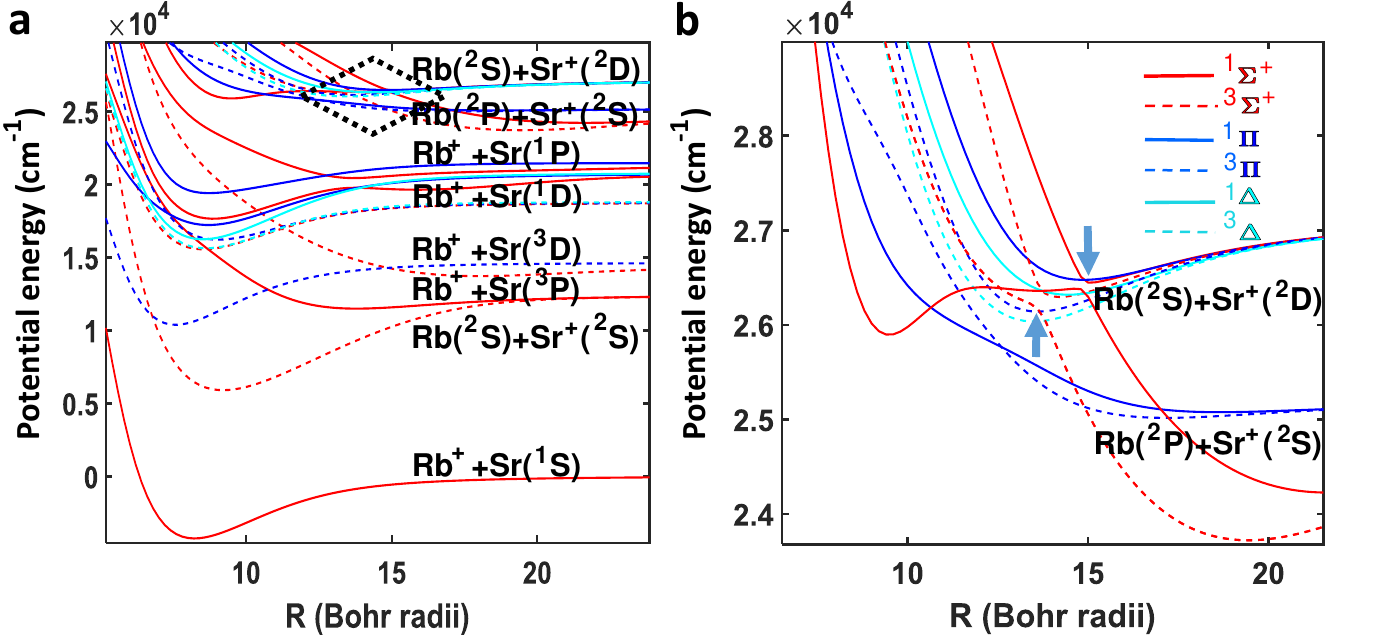}
		\caption[molecular potentials2]{\footnotesize{\textbf{Ab-initio calculated molecular potential energy curves for the RbSr$^{+}$ molecular ion.} \textbf{(a)} PECs up to the eighth dissociation limit in the Hund's case \textit{a} (no spin-orbit coupling) representation. The entrance channel of our system is the upper Rb($5s\,^2S$)+Sr$^{+}$($4d\,^2D$) asymptote. \textbf{(b)} Zoom in on  the diamond-shaped box in panel $\textbf{a}$, showing the two asymptotes relevant for our experiment. An avoided crossing is clearly visible in the $^1\Sigma^+$  and $^3\Sigma^+$ symmetries, indicated by the two arrows, located at $R_1=14.9\,a_0$ and $R_3=13.7\,a_0$ correspondingly, where $a_0$ is the Bohr radius.}}
		\label{adiabatic2}
	\end{figure} 
	Considering such well-defined transitions to levels with lifetimes that extend beyond the typical Langevin collision time, one could expect the width of the histograms to be as narrow as dictated by the finite lifetime of the levels involved. We believe that the excessive width of the energy distributions we measured is due to inaccuracies in our SSDCT measurements, rather than to an actual physical broadening mechanism. Previously SSDCT was shown to be accurate to within 5$\%$ for energies up to 250 K$\cdot$k$_B$ \cite{single-shot}, given a well known initial distribution of the energy between the trap modes. But for energies as high as 1500 K$\cdot$k$_B$ we expect larger inaccuracies, in particular in the way different directions of motion are analyzed (see Supplementary Material and \cite{single-shot, Direct-sympathetic}). The broad energy distribution also prevents the observation of spectral separation of 340 K$\cdot$k$_B$ between the two EEE channels to the two fine-structure split Rb($P_{1/2,3/2}$) levels. 
	
	It is worth mentioning, that we did not observe events in which the ion kinetic energy was above 10000 K$\cdot$k$_B$. This indicates that there is no channel in which the ion quenches directly from Rb($^2S$)+Sr$^{+}$($^2D$) to the Rb($^2S$)+Sr$^{+}$($^2S$) which would have released $\sim$ 21000 K$\cdot$k$_B$ into kinetic motion. We substantiated this argument further by lowering the ion's trap depth to few 1000's of K$\cdot$k$_B$ and did not observed any ion loss due to extensive heating.  
	
	A similar analysis of the energies measured using SSDCT when the ion is initialized in the $D_{3/2}$ state is shown by the lower red histogram in Fig.~\ref{Hist6}a. As seen, here only the $\simeq$ 1500 K$\cdot$k$_B$ EEE channel remains. The SOC channel cannot be endothermically excited at such a low initial temperature. Therefore, the SOC channel can be turned on or off by initializing the ion in the $D_{5/2}$ or $D_{3/2}$ level.
	
	Finally, we compared the spectrum of energies released between the cases where the electronic spins of the ion and atoms are aligned parallel or anti-parallel (same levels as those used in Fig.~\ref{quench}b). The two spectra are shown in Fig.~\ref{Hist6}c. As seen, no significant difference can be observed between the two cases, again suggesting strong angular-momentum mixing. 
	
	We now turn to a theoretical analysis of the quench process. Figure~\ref{Hist6}d provides a schematic illustration of the dynamics we observed. When the ground state Rb atom and the excited Sr$^+$ ion approach each other their electronic wavefunctions are strongly perturbed by their interaction, leading to avoided crossings between molecular potential curves. Non-adiabatic transition through these crossings leads to the large release of kinetic energy.
	
	\begin{figure}[h!]
		\includegraphics[width=0.5\textwidth]{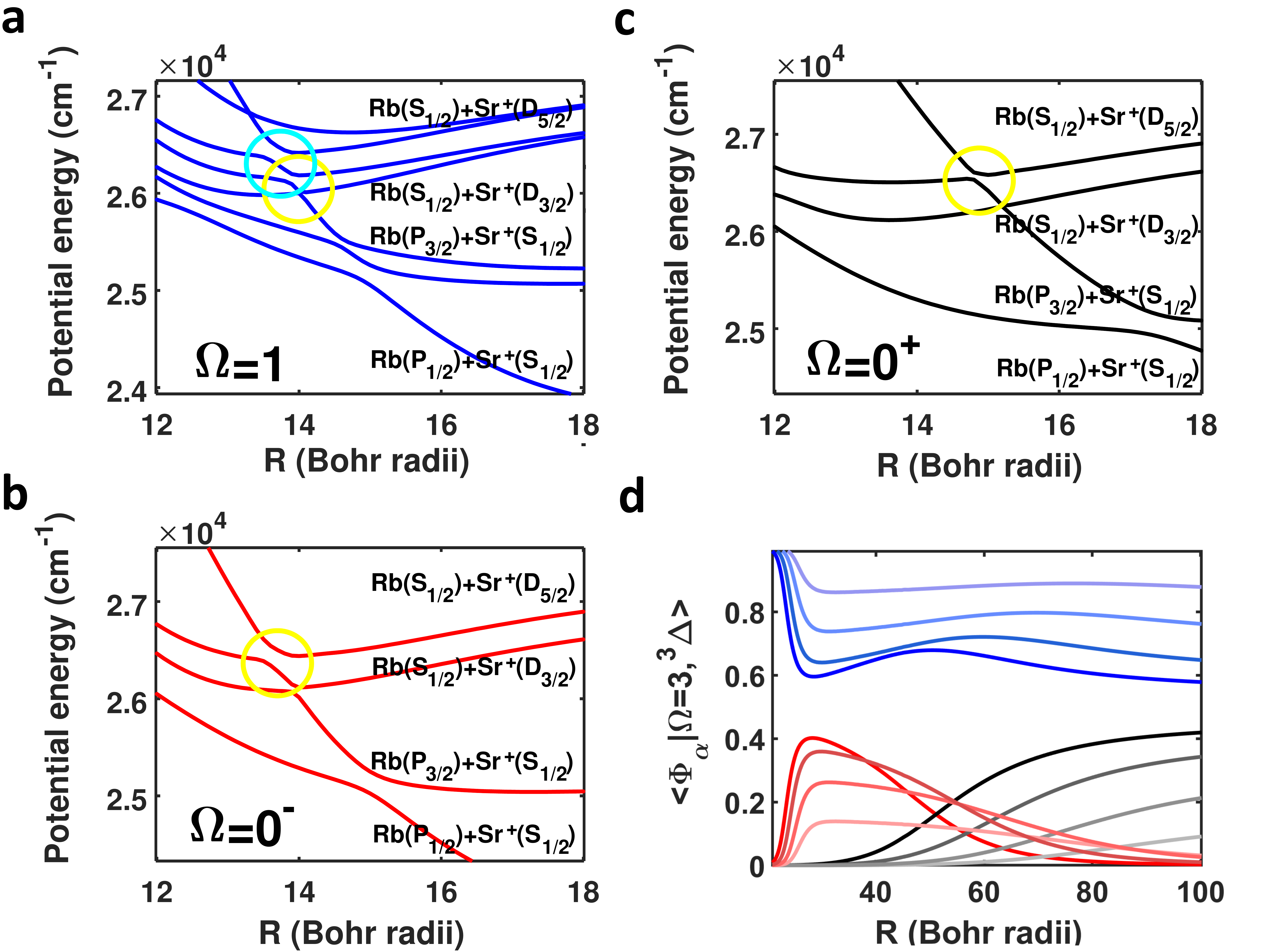}
		\caption[molecular potentials]{\footnotesize{\textbf{Potential energy curves for the excited states including spin-orbit interaction.} \textbf{(a,b,c)} RbSr$^+$ PECs in Hund's case \textit{c} representation, including SO interaction for the two displayed asymptotes (see text). The symmetries $\Omega=1,0^{\pm}$ are displayed as blue, red, black lines in different frames for clarity. The avoided crossings for the SOC (cyan) and EEE (yellow) processes are marked with circles.} \textbf{(d)} The component on the $\Omega=3$, $(^3\Delta)$ Hund's case \textit{c} state of three of the eigenvectors $\Phi_\alpha$ (red, blue and black curves) of the full Hamiltonian matrix including SOC and Coriolis coupling (see Supplementary Material). Each is plotted for four different values of the maximal partial wave, J=3, 5, 10, 20 (from bright to dark color, respectively).}  
		\label{diabatic7}
	\end{figure} 
	
	Figure~\ref{adiabatic2}a presents the numerically calculated RbSr$^+$ PECs without spin-orbit interaction (Hund's case \textit{a}) \cite{aymar2011}. The collision entrance channel considered here is the eighth dissociation limit of the RbSr$^+$ molecular ion. Figure~\ref{adiabatic2}b presents a closer view of this channel. As seen, two molecular symmetries, $^1\Sigma^+$ and $^3\Sigma^+$, undergo an avoided crossing to a different outgoing channel. By linearizing these PECs around the avoided crossing, we calculated Landau-Zener (LZ) transition probabilities $P(^1\Sigma^+(i\rightarrow f))=0.39$ and $P(^3\Sigma^+(i\rightarrow f))=0.46$ (see Supplementary Material for details). Taking into account the statistical population of these symmetries assuming equal distribution over all possible symmetries, reduces the expected tunneling probability to the few percent level, well below our observed rates. Therefore, while the Hunds case \textit{a} analysis qualitatively explains our observations it does not quantitatively predict our results. Figure~\ref{adiabatic2}a also explains why charge-exchange was not observed as the collision entrance channel has no avoided crossings with Rb$^+$+Sr PECs which are situated well below in energy. 
	
	In order to explain the SOC channel and qualitatively improve our analysis, we included spin-orbit (SO) coupling (Hund's case \textit{c}). As the molecular $R$-dependent SO coupling is unknown, we rely on the atomic SO \cite{beuc2007,LozeilleEPJD2006}. This is reasonable since the avoided crossings are located at distances comparable to the estimated range of the exchange interaction. In order to take into account the change to atomic wavefunction across the LZ crossings we used the linearized form of the $^1\Sigma^+$ and $^3\Sigma^+$ symmetries (see Supplementary Material). 
	Here, we observed avoided crossings in three of the resulting symmetries $\Omega=0^\pm, 1$, shown in Fig.~\ref{diabatic7}. No avoided crossings occur in the $\Omega=2,3$ symmetries. We identified EEE (SOC) transitions through three (one) avoided crossings marked by yellow (cyan) circles. Similarly to  Hund's case \textit{a} we analyzed the expected tunneling probabilities through these avoided crossings. All calculated and measured tunneling probabilities are compared in Table~\ref{table:prob}, together with the expected and measured rate-ratios between the different EEE and SOC channels. The last column refers to a maximal LZ probability of 0.5 at all avoided crossings (For more details see Supplementary Material). 
	\begin{table}[h!]
		\begin{center}
			\begin{tabular}{ ||p{2cm}||m{2cm}m{2cm}m{2cm}||}
				\hline
				\multicolumn{4}{||c||}{\bf{Summary of the transition probabilities}.} \\
				\hline
				&&&\\
				&$Experiment$&$Theory$&$Theory$\\
				&&&$P_{LZ}=0.5$\\
				\hline
				$P^{5/2}_{EEE+SOC}$&0.34$\pm{0.017}$&0.2&0.2\\
				$P^{3/2}_{EEE}$&0.45$\pm{0.018}$&0.03 &0.125\\
				$P^{5/2}_{EEE}/P^{3/2}_{EEE}$&0.53$\pm{0.043}$&6&1.28\\
				$P^{5/2}_{SOC}/P^{5/2}_{EEE}$&0.39$\pm{0.05}$&0.11&0.25\\
				\hline
			\end{tabular}
			\caption{The probabilities for the various channels extracted from experiment and from theory under two different hypothesis. The first and second rows are the measured and calculated total probabilities to quench from the $D_{5/2}$ ($P^{5/2}_{EEE+SOC}$) and $D_{3/2}$ ($P^{3/2}_{EEE}$), respectively. The last two rows are the probability-ratios between the different EEE and SOC channels.}
			\label{table:prob}
		\end{center}
	\end{table}
	As seen in the table, while qualitatively producing the EEE and SOC channels, our experimentally measured probabilities are higher than the predictions of theory. This disagreement persists even when considering the maximal transition probability case. Furthermore, as shown above, even when preparing the atom-ion complex with total angular-momentum 3$\hbar$ the quenching probability remains high, while conservation of internal angular-momentum implies complete protection against non-adiabatic quench. Both these observations suggest significant coupling between the Hunds case \textit{c} PECs and transfer of internal electronic angular-momentum to molecular rotation.
	
	The transfer between internal angular-momentum and molecular rotation occurs through inertial coupling. Typically, in ultracold collisions, such inertial coupling is small due to the small number of partial waves involved. However, in atom-ion collisions the number of partial waves can be large even in the few 100's $\mu$K$\cdot$k$_B$ energy range, due to the relatively long-ranged $R^{-4}$ atom-ion interaction. In our experiment up to 20 partial waves contribute in the collision, and therefore molecular rotation  may be strongly coupled to internal electronic angular-momentum. This results in coupling between the PEC's in Fig.~\ref{diabatic7} and therefore the hypothesis of population distributed statistically across all PEC's, used in deriving the theoretical probabilities in Table~\ref{table:prob} is compromised.
	
	To show the effect of inertial coupling, we add a rotational term to the SO matrices. Strong angular-momentum mixing is confirmed by the results of Fig.~\ref{diabatic7}d. In the molecular structure calculations, the fully-aligned situation is correlated to the molecular state with the highest angular-momentum, namely the $\Omega=3$ Hund's case \textit{c} state, composed only by the $^3\Delta$ Hund's case \textit{a} state. This symmetry does not exhibit any avoided crossing, so that it must be strongly mixed with other molecular state to undergo a quenching process. As an illustration, Fig.~\ref{diabatic7}d. displays the component on the $\Omega=3$, $(^3\Delta)$ Hund's case \textit{c} state of three of the eigenvectors $\Phi_\alpha$ of the full Hamiltonian matrix including SOC and Coriolis coupling. Each is also plotted for four values of the maximal partial wave, J=3, 5, 10, 20. These plots demonstrate that the $\Omega=3$, $(^3\Delta)$ is indeed strongly coupled to other states at infinity, and the coupling increases with J value. In case of s-wave collision, where J=3, still exists a degree of angular mixing, indicating that this fully aligned state lacks protection.  In other words, none of the internal angular-momentum of the colliding partners are conserved when moving to the molecular frame. \\
	
	To conclude, we have studied the dynamics of an ion, initialized in a metastable excited electronic level, during collisions with ultracold atoms. By measuring the final kinetic energy of the ion after a single collision we found that two non-adiabatic quench processes occur. The first is an excitation-exchange between the atom and ion and the second is a change of the fine-structure level of the ion excited state. These non-radiative decay channels happen through an avoided crossing, releasing the internal energy into molecular motion. Comparing our results with molecular structure calculations suggests that due to the high partial-waves involved in this collision, the electronic angular-momentum is mixed through Coriolis coupling, leading to the transfer of electronic angular-momentum to external nuclei rotation. Our findings shed light on the dynamics of inelastic atom-ion collisions and pave the way for controlling cold chemical reactions at a single collision level.
	\paragraph{Acknowledgments---}
	This work was supported by the Crown Photonics Center, the Israeli Science Foundation, the Israeli Ministry of Science Technology and Space and the European Research Council (consolidator grant 616919-Ionology)
	
	\onecolumngrid
	\newpage
	\begin{center}
		\Large{\textbf{Supplemental Material}}
	\end{center}
	\subsection{Single-Shot Doppler Cooling Thermometry (SSDCT)}
	In order to detect the energy spectrum of the ion after a quench event, we use the numerical simulation for the energy, and fluoresence dynamics during laser cooling described in \cite{single-shot}. Since here the ion reached higher energies as compared with \cite{single-shot}, longer Doppler cooling times were used due to larger Doppler shifts (up to 25 s of cooling times). Our simulation calculates the ion's trajectory up to 250 K$\cdot$ k$_B$. Therefore the lower energy histogram distributed around 300~ms and attributed to an energy of 200 K$\cdot$ k$_B$ should be reasonably accurate (the accuracy of this method was estimated to be on the level of 5$\%$ if the direction of motion of the ion is known). In order to model higher energies, we extrapolated a fit $a\*(x_0-x)^{n}$ from the numerical simulation, were $x$ is the cooling time, and $x_0$ is the initial detection time. By fixing the $a$ and $n$ parameters from the fit, we translate all rising times into the energies of the ion. We assume that the ion is projected isotropically in space after a collision, therefore the curve was simulated for initial radial energy distribution (250 K$\cdot$ k$_B$ on the $x$ radial axis). The differences between the two radial modes were found to be in the few Kelvin regime.
	\subsection{Collisional Quenching Rates}
	In order to measure quenching rate from the $D_{5/2}$ state, we initially prepared the ion in the $D_{5/2}$ state by applying a shelving pulse on the $5s^2S_{1/2} \rightarrow 4d^2D_{5/2}$ quadrapole transition with a narrow-linewidth 674~nm laser. After a collision, we apply the Doppler cooling transition $5s^2S_{1/2} \rightarrow 5p^2P_{1/2}$, indicating whether population either quenched to the $S$ state or remained in $D_{5/2}$ state (referred to as ''bright'' or ''dark'' events). The $D_{3/2}$ state is prepared by using a laser pulse on the cooling transition. After a collision, we shelve population to the $D_{5/2}$ state by a laser pulse on the $5s^2S_{1/2} \rightarrow 5p^2P_{3/2}$ dipole transition, followed by detection. In contrast to the previous case, ''bright'' and ''dark'' characters mean that the ion remained or quenched from the $D_{3/2}$ state, respectively. Since the $P_{3/2}$ state can decay both to the $D_{5/2}$ ``dark'' state and the $D_{3/2}$ ``bright'' state, 11$\%$ of the quenched events in the $S$ are detected as non-quenched ``bright'' events because they are pumped to the wrong state in our detection. For that, the $D_{3/2}$ state exponential curve decays to the branching ratio constant (see Sec.\ref{Rate Equation}).
	
	To overlap the atomic cloud with the ion, we move the atoms from a position of 50 $\mu m$ above the ion in 5 ms using a piezo-driven mirror which controls the position of the atomic optical dipole trap. Since the Rb cloud has a Gaussian profile, the ion starts to interact 0.5 ms before the dipole trap reaches zero position. This time offset is added to all interaction times (x-axis). 
	First point for the $D_{5/2}$ data indicates loss of fidelity in the shelving pulse during the initialization of the ion in $D_{5/2}$ state. This error originates from drifts in the transition frequency.
	\subsection{Rate Equation}
	\label{Rate Equation}
	The collisional quenching rate $\Gamma_{Q}$ is extracted from the solution of the rate equation: 
	\begin{equation}
	\dot{P_{D}}=-(\Gamma_{Q}+\Gamma_{OP})(P_{D}-c),
	\label{rate}
	\end{equation}
	where $P_{D}$ and $\Gamma_{OP}$ are the $D$ state population (normalized to one at the moment of the preparation) and the ion decay rate in the absence of atoms, respectively, and $c$ is an offset constant. The collisions occur in the presence of the 1064 nm ODT. Therefore, $\Gamma_{OP}$ accounts for the spontaneous decay rate, and for the optical-pumping rate caused by the 1064 nm photon scattering which off-resonantly couples the $D$ state back to the $S$ state, through the $P$ state. The initial population $P_{D}(t=0)$ and $\Gamma_{OP}$ are extracted from the fit of the ion decay curve without atoms, and set as fixed parameters. Setting $c=0$, we find a lifetime $\tau_{OP} \equiv \Gamma_{OP}^{-1}=58.6(3.2)$ ms for the $D_{5/2}$ level, significantly shorter than the natural lifetime of 390 ms. In the presence of atoms the Langevin collisions occur at a rate $1/\tau_L=2\pi n\sqrt{e^{2}\alpha/\mu}$,  where $n$ is the density of the Rb cloud. $\tau_L$, $e$, $\alpha$ and $\mu$ are Langevin time constant, electron charge, Rb atom static dipole polarizability and the RbSr$^+$ reduced mass, respectively. The ion quickly decays after $\Gamma_{Q}=2.6(4)$ Langevin collisions on average (Fig.~\ref{quench}.a, blue curve). 
	$\Gamma_{Q}$ is thus the only free parameter, and $c$ is set to zero for the $D_{5/2}$ case. 
	
	We repeated the same measurement when the ion is prepared in the $D_{3/2}$ level. As a result of our detection scheme, we measured the branching ratio $c=0.11(1)$ between the two dipole-allowed transition rates $\Gamma_{P_{3/2}\rightarrow D_{5/2}}$ ($=1$ MHz) and $\Gamma_{P_{3/2}\rightarrow D_{3/2}}$ ($=8.7$ MHz) \cite{NIST}, which fixes the value of the offset parameter in Eq.~\ref{rate}. We found that without the presence of atoms, the lifetime of the $D_{3/2}$ level is $\tau_{OP}= 45.2(2.1)$~ms. In the presence of atoms the ion decays after $\Gamma_{Q}=2.8(3)$ Langevin collisions on average.
	\subsection{Landau-Zener Probabilities}
	Out of the molecular complexity shown in Fig.~\ref{adiabatic2}, 
	a great simplification occurs. The observed reaction is mainly attributed to two avoided crossings occurring in two molecular symmetries. We use the classical Landau-Zener model \cite{landau1932}, to find the probability $0 \le P(i\rightarrow f) \le 0.5$ for the atom-ion pair colliding in the entrance channel ''$i$" with an initial relative energy $E_i$ to exit in the final channel ''$f$'',
	\begin{equation}
	P(i\rightarrow f; E_i)=2P^{LZ}(i;E_i)(1-P^{LZ}(i;E_i)),
	\label{pass}
	\end{equation}   
	where $P^{LZ}(i;E_i)$ is the probability for a single-path through the crossing: 
	\begin{equation}
	P^{LZ}(i;E_i)=\exp{\frac{-2\pi W^{2}_{c} }{(v_{c}\Delta F_{c})}}. 
	\label{landau}
	\end{equation}   	
	This expression assumes that the two involved PECs are separated by an energy $2W_c$ at the crossing point $R_c$ defined by symmetrically linearizing the PECs around the avoided crossing. This results in two local lines with slopes $F_i$ and $F_f$ (with $\Delta F_{c}=|F_i-F_f|$) crossing at $R_c$. The relative collision velocity $v_c$ at the crossing point is defined by $v_c=\sqrt{2\mu(E_i-U_c)}$, where $U_c$ is the potential energy at the crossing point. In practice, when an avoided crossing is well localized, as in the present case, a reasonably accurate linearization can be achieved. Note that at ultracold energies, $E_i$ is negligible against $U_c$ and has been omitted for simplicity. 
	This leads to single-path probabilities $P^{LZ}(^1\Sigma^+)=0.73$ and $P^{LZ}(^1\Sigma^+)=0.36$, and to the final probabilities reported in the paper, $P(^1\Sigma^+(i\rightarrow f))=0.39$ and $P(^3\Sigma^+(i\rightarrow f))=0.46$.
	These probabilities correspond to an initial full population in the considered symmetry, and such avoided crossings are quite efficient for the reaction to occur. But assuming a statistical population of these molecular symmetries when entering the Rb($5^2S$)+Sr$^+$ ($4^2D$) channel, a statistical weight of 1/20 for $^1\Sigma^+$ and 3/20 for $^3\Sigma^+$ must multiply these probabilities, reducing them to a few percent, well below the observed rates.
	
	Now switching to Hund's case \textit{c} (see Sec.\ref{Spin-Orbit}), let us first consider, referring from Fig.~\ref{diabatic7}, 
	the $0^{+/-}(D_{5/2})$ PECs of the entrance channel Rb($S_{1/2}$)+Sr$^+$($D_{5/2}$). First assuming a full population in each of the symmetries, we find for the avoided crossings responsible for EEE $P^{LZ}(0^+(D_{5/2}))=0.88$ and $P^{LZ}(0^-(D_{5/2}))=0.52$, and  $P(0^+(D_{5/2})(i\rightarrow f))=0.21$ and $P(0^-(D_{5/2})(i\rightarrow f))=0.49$. The introduction of the SO coupling tends to weaken the efficiency of the $0^+$ avoided crossing compared to the $^1\Sigma^+$ case, while the $0^-$ avoided crossing has now an almost maximal transition probability. Considering now a statistical population of the states in the entrance channel, the weight of $0^+(D_{5/2})$ and $0^-(D_{5/2})$ is 1/12, yielding to probabilities of $P(0^+(D_{5/2})(i\rightarrow f))=0.02$ and $P(0^-(D_{5/2})(i\rightarrow f))=0.04$, respectively.
	
	The case of $\Omega=1$ is a slightly more complicated as two successive avoided crossings are identified. The crossing circled in cyan involves the lower $\Omega=1(D_{5/2})(1)$ and the upper $\Omega=1(D_{3/2})(2)$ PECs, and  the one circled in yellow the $\Omega=1(D_{3/2})(2)$ and the $\Omega=1(D_{3/2})(1)$ PECs. The single-path probability associated to each of them is ${\cal P}\equiv P^{LZ}(\Omega=1(D_{5/2})(1))=0.59$, and ${\cal P'}\equiv P^{LZ}(\Omega=1(D_{3/2})(2))=0.86$, respectively. To deduce the EEE and SOC probabilities, we have to combine these values as $P^{SOC}(\Omega=1(D_{5/2})(1))(i\rightarrow f))={\cal P}(2-{\cal P})(1-{\cal P'})=0.11$, and $P^{EEE}(\Omega=1(D_{5/2})(1))(i\rightarrow f))={\cal P}(2-{\cal P}){\cal P'}=0.71$. With the statistical weight of 2/12 for the $\Omega=1(D_{5/2})(1)$ state, and summing with the EEE probabilities for the $0^+(D_{5/2})$ and $0^-(D_{5/2})$ symmetries, we obtain a total probability for EEE starting from the Rb($S_{1/2}$)+Sr$^+$($D_{5/2}$) channel of $P^{EEE}(D_{5/2})=0.18$, and for SOC of $P^{SOC}(D_{5/2})=0.02$. Similarly, one can determine the total EEE probability $P^{EEE}(D_{3/2})=2{\cal P'}(1-{\cal P'})/4=0.03$, where we assigned a statistical weight of 1/4 to the $\Omega=1(D_{3/2})(2)$ state.
	
	At this stage it is worthwhile estimating the accuracy of the proposed simple LZ model. As quoted above, the double-path transition probability $P(i\rightarrow f; E_i)$ yielded by the LZ model is at most equal to 0.5, obtained for a single-path probability $P^{LZ}(i;E_i)=0.5$. With this maximal value, and keeping the same statistical weights, one would obtain $P(0^+(D_{5/2})(i\rightarrow f))=P(0^-(D_{5/2})(i\rightarrow f))=0.5/12=0.04$. Similarly, one would get $P^{SOC}(\Omega=1(D_{5/2})(1))(i\rightarrow f))=P^{EEE}(\Omega=1(D_{5/2})(1))(i\rightarrow f)=0.375/6=0.06$. In total one would reach the probabilities $P^{EEE}(D_{5/2})=0.14$, $P^{SOC}(D_{5/2})=0.06$, and $P^{EEE}(D_{3/2})=0.125$.
	
	\subsection{Spin-Orbit Coupling}
	\label{Spin-Orbit}
	The PECs in Fig.~\ref{adiabatic2} do not include spin-orbit interactions. Such couplings mix Hund's case  \textit{a} PECs, and is of particular importance when they exhibit real or avoided crossings. Here we assume a simple atomic model to describe the spin-orbit interaction. The Hamiltonian matrices involve the interaction parameter $A_{sp}=\frac{2}{3}\Delta E_{P}$, and $A_{sd}=\frac{2}{5}\Delta E_{sd}$, where $\Delta E_{sp}$=237.595 cm$^{-1}$ and $\Delta E_{SD}$=280.34 cm$^{-1}$ are the atomic spin-orbit energy splittings of Rb($5^2P$) and Sr$^+$($4^2D$), respectively.
	The Hamiltonian matrices coupling the PECs correlated to the Rb($5^2P$)+Sr$^+$($5^2S$) dissociation limit, are expressed for each possible value of the projection $\Omega=0^{\pm} ,1,2$ of the total electronic angular momentum as follows \cite{beuc2007}:	
	\begin{gather}	
	H_{\Omega=2}=V(^{3}\Pi)+\frac{A_{sp}}{2},
	\label{4}
	\end{gather}
	\begin{gather}
	H_{\Omega=1}= 
	\begin{bmatrix}
	V(^{3}\Pi)&-\frac{A_{sp}}{2} & \frac{A_{sp}}{2} \\
	-\frac{A_{sp}}{2}&V(^{1}\Pi)&\frac{A_{sp}}{2} \\
	\frac{A_{sp}}{2} &\frac{A_{sp}}{2} & V'(^{3}\Sigma^{+})\\                  
	\end{bmatrix},
	\end{gather}
	\begin{gather}
	H_{\Omega=0^{-}}=
	\begin{bmatrix}
	V(^{3}\Pi)-\frac{A_{sp}}{2}&-\frac{A_{sp}}{\sqrt{2}}\\
	-\frac{A_{sp}}{\sqrt{2}}
	& V'(^{3}\Sigma^{+})                
	\end{bmatrix},
	\end{gather}
	\begin{gather}
	H_{\Omega=0^{+}}=
	\begin{bmatrix}
	V(^{3}\Pi)-\frac{A_{sp}}{2}&-\frac{A_{sp}}{\sqrt{2}}\\
	-\frac{A_{sp}}{\sqrt{2}}
	& V'(^{1}\Sigma^{+})                
	\end{bmatrix},
	\end{gather}  	
	Similarly, the Hamiltonian matrices (with $\Omega=0^{\pm} ,1,2,3$) coupling the PECs correlated to the Rb($5^2S$)+Sr$^+$($4^2D$) dissociation limit are expressed as follows \cite{LozeilleEPJD2006}:	
	\begin{gather}	
	H_{\Omega=3}=V(^{3}\Delta)+A_{sd},
	\end{gather}
	\begin{gather}
	H_{\Omega=2}=
	\begin{bmatrix}
	V(^{3}\Delta)&A_{sd}& \frac{A_{sd}}{\sqrt{2}}\\
	A_{sd}&V(^{1}\Delta)&-\frac{A_{sd}}{\sqrt{2}}\\
	\frac{A_{sd}}{\sqrt{2}}&-\frac{A_{sd}}{\sqrt{2}}& V(^{3}\Pi)+\frac{A_{sd}}{2}                  
	\end{bmatrix},
	\end{gather}
	\begin{gather} 	 	 
	H_{\Omega=1}=
	\begin{bmatrix}
	V(^{3}\Delta)-A_{sd}&-\frac{A_{sd}}{\sqrt{2}}&\frac{A_{sd}}{\sqrt{2}}&0\\
	-\frac{A_{sd}}{\sqrt{2}}& V(^{3}\Pi)&-\frac{A_{sd}}{2}&\frac{A_{sd}\sqrt{3}}{2}\\
	\frac{A_{sd}}{\sqrt{2}}&-\frac{A_{sd}}{2} & V(^{1}\Pi)&\frac{A_{sd}\sqrt{3}}{2} \\
	0&  \frac{A_{sd}\sqrt{3}}{2}&\frac{A_{sd}\sqrt{3}}{2}& V'(^{3}\Sigma^{+})           
	\end{bmatrix},
	\end{gather}
	\begin{gather} 	 	 
	H_{\Omega=0^{-}}=
	\begin{bmatrix}
	V(^{3}\Pi)-\frac{A_{sd}}{2}&A_{sd}\sqrt{\frac{3}{2}}\\
	A_{sd}\sqrt{\frac{3}{2}}
	& V'(^{3}\Sigma^{+})                
	\end{bmatrix},
	\end{gather} 
	\begin{gather}	 	 
	H_{\Omega=0^{+}}=
	\begin{bmatrix}
	V(^{3}\Pi)-\frac{A_{sd}}{2}&-A_{sd}\sqrt{\frac{3}{2}}\\
	-A_{sd}\sqrt{\frac{3}{2}}
	& V'(^{1}\Sigma^{+})               
	\end{bmatrix}.
	\label{12}
	\end{gather}
	The $V(\Lambda)$ are the PECs plotted in Fig.~\ref{adiabatic2}, except from the $^1\Sigma^+$ and $^3\Sigma^+$ symmetries for which we used the curves linearized around their avoided crossing (hereafter referred to as $V'(^{1}\Sigma^{+})$ and $V'(^{3}\Sigma^{+})$). In these latter cases, we model the interaction around the crossing point $R_c^j$ with a Gaussian form (the index $j=1,2$ corresponding to the $^1\Sigma^+$ and $^3\Sigma^+$ symmetries, respectively)
	\begin{equation}
	G^{(j)}=2W_c^{j}\exp^{-(\frac{R-R_c^j}{\delta R^{j}})^{2}},
	\end{equation}
	where $2W_c^{j}$ is the energy gap at $R_c^j$ between the Hund's case \textit{a} original PECs (see main text), and $\delta R^{j}$ the estimated width of the avoided crossing. The values of these parameters for $^1\Sigma^+$($^3\Sigma^+$) symmetries are $2W_c^{j}=2.51\cdot10^{-4}\   (5.74\cdot10^{-4})$ a.u., $R_c^j$ = 14.9 (13.7) $a_{0}$, $\delta R^{j}=0.15$ (0.3) $a_{0}$,  respectively.
	The PECs displayed in Fig.~\ref{diabatic7}.a,b,c for $\Omega=0^{\pm} ,1$ ($\Omega=2,3$ are not shown) thus result from the diagonalization of the following matrices which include previous matrices for the Rb($5^2P$)+Sr$^+$ ($5^2S$) and Rb($5^2S$)+Sr$^+$ ($4^2D$) subspaces, connected by the Gaussian coupling terms:
	\begin{align} 	 	 
	H_{\Omega=0^{-}}=
	\begin{bmatrix}
	V(^{3}\Pi)-\frac{A_{sd}}{2}&-A_{sd}\sqrt{\frac{3}{2}}&0&0\\
	-A_{sd}\sqrt{\frac{3}{2}}& V'(^{3}\Sigma^{+})&0&G^{(2)}\\
	0&0&V(^{3}\Pi)-\frac{A_{sp}}{2}&-\frac{A_{sp}}{\sqrt{2}}\\
	0&G^{(2)}&-\frac{A_{sp}}{\sqrt{2}}&V'(^{3}\Sigma^{+})
	\end{bmatrix},
	\end{align} 
	\begin{align} 	 	 
	H_{\Omega=0^{+}}=
	\begin{bmatrix}
	V(^{3}\Pi)-\frac{A_{sd}}{2}&-A_{sd}\sqrt{\frac{3}{2}}&0&0\\
	-A_{sd}\sqrt{\frac{3}{2}}&V'(^{1}\Sigma^{+})&0&G^{(1)}\\      
	0&0&V(^{3}\Pi)-\frac{A_{sp}}{2}&-\frac{A_{sp}}{\sqrt{2}}\\
	0& G^{(1)}&-\frac{A_{sp}}{\sqrt{2}}& V'(^{1}\Sigma^+)
	\end{bmatrix},
	\end{align} 
	The $\Omega=1$ matrix is built in the same manner, assembling the two $\Omega=1$ blocks for Rb($5^2P$)+Sr$^+$($5^2S$) and Rb($5^2S$)+Sr$^+$($4^2D$), and adding the $G^{(2)}$ term.
	\subsection{Coriolis Coupling}
\begin{figure}[h!]
		\includegraphics[width=0.65\textwidth]{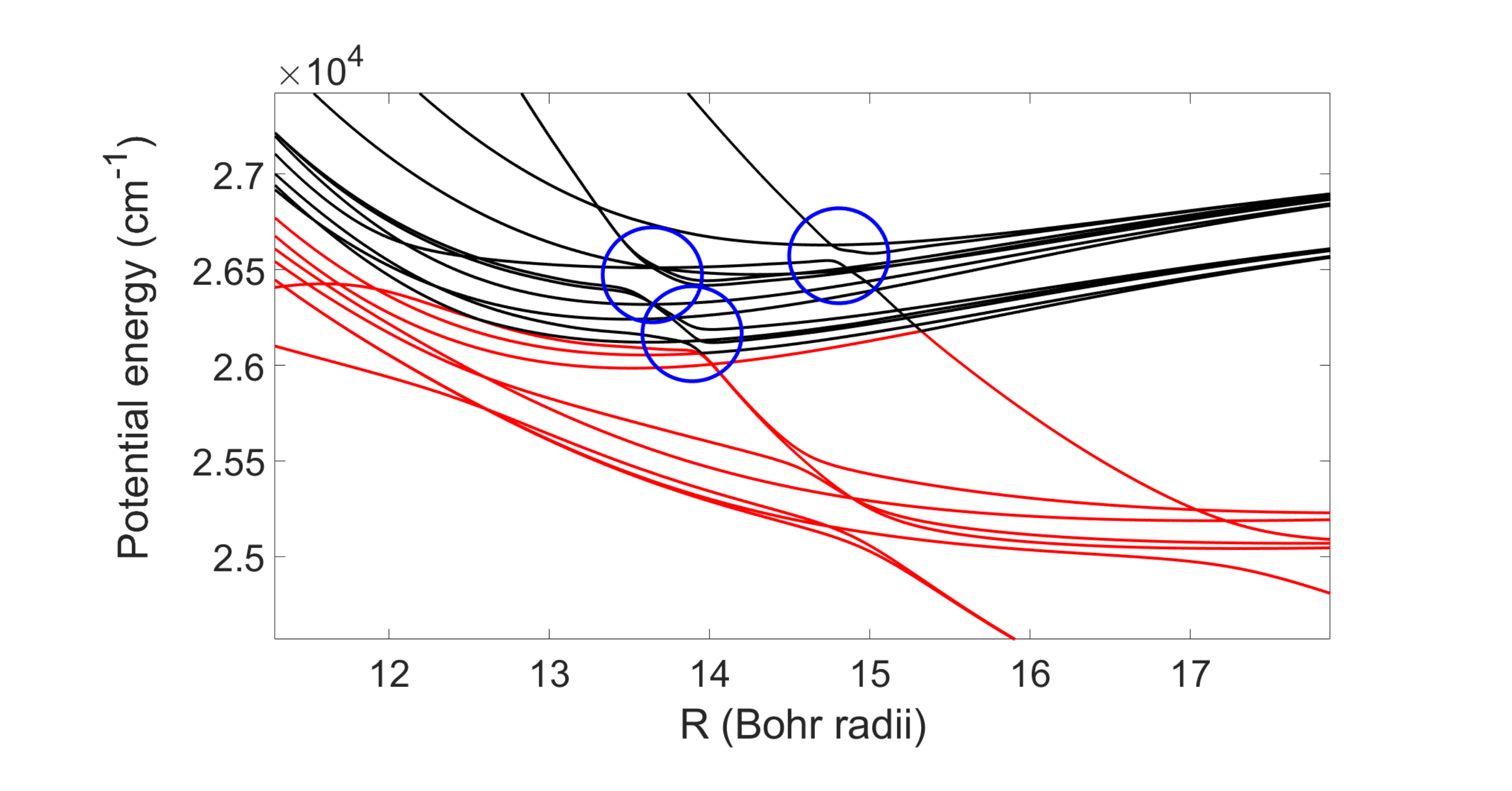}
		\caption[molecular potentials3]{\footnotesize{\textbf{The eigenenergies obtained after the diagonalization of the full matrix.} PECs in Hund's case \textit{c} representation, including spin-orbit, Coriolis and Gaussian couplings. The blue circled crossings originate from the crossings from Fig.~\ref{diabatic7}. More crossings are manifested indicating an even stronger mixing deriving from the spin-Coriolis and orbital angular momentum-Coriolis which highly couple the states.}}
		\label{eigenenergy_CC}
	\end{figure}

	Here we describe the full coupling matrix which includes SO and Coriolis coupling terms, the latter mixing states with different $\Omega$ symmetries. The plots in Fig.~\ref{diabatic7}.d and the potentials energy plotted in Fig.~\ref{eigenenergy_CC}, result from the diagonalization of this matrix.
	First we introduce few short-hand notations defining the matrix elements. The diagonal and off-diagonal parts of the full Hamiltonian $H^{^{2S+1}|\Lambda|}$ include the angular, Coriolis coupling terms, where the full description can be found in \cite{Moszynski2006,Moszynski2012}:\\  
	\begin{equation*}
	H_{diag}^{^{2S+1}|\Lambda|}=V^{^{2S+1}|\Lambda|}(R)+\frac{J(J+1)+S(S+1)+L(L+1)-\Omega^{2}-\Sigma^{2}-\Lambda^{2}}{2\mu R^{2}},
	\end{equation*}
	\begin{equation}
	H_{offdiag}^{^{2S+1}|\Lambda|}=\frac{(J^{+}S^{-}+J^{-}S^{+})+(L^{+}S^{-}+L^{-}S^{+})-(J^{+}L^{-}+J^{-}L^{+})}{2\mu R^{2}},
	\end{equation} 
	where we use the operators:
	\begin{equation}
	S^{2}\Phi^{S\Sigma\Lambda}_{n\Omega}=S(S+1)\Phi^{S\Sigma\Lambda}_{n\Omega},\\
	\end{equation}
	\begin{equation}
	S_{z}\Phi^{S\Sigma\Lambda}_{n\Omega}=\Sigma\Phi^{S\Sigma\Lambda}_{n\Omega},\\
	\end{equation}
	\begin{equation}
	L_{z}\Phi^{S\Sigma\Lambda}_{n\Omega}=\Lambda\Phi^{S\Sigma\Lambda}_{n\Omega},\\
	\end{equation}
	\begin{equation}
	S^{\pm}\Phi^{S\Sigma\Lambda}_{n\Omega}=[S(S+1)-\Sigma(\Sigma\pm1)]^{1/2}\Phi^{S\Sigma\pm1\Lambda}_{n\Omega\pm1},\\
	\end{equation}
	\begin{equation}
	L^{\pm}\Phi^{S\Sigma\Lambda}_{n\Omega}=[L(L+1)-\Lambda(\Lambda\pm1)]^{1/2}\Phi^{S\Sigma\Lambda\pm1}_{n\Omega\pm1},\\
	\end{equation}
	\begin{equation}
	J^{\pm}\Phi^{S\Sigma\Lambda}_{n\Omega}=[J(J+1)-\Omega(\Omega\pm1)]^{1/2}\Phi^{S\Sigma\Lambda}_{n\Omega\pm1},
	\end{equation} 
	which acts on the electronic wave function $\Phi^{S\Sigma\Lambda}_{n\Omega}(r,R)$.\\
	For simplicity, we use short-hand notations describing the rotational couplings between $J$, $S$, and $L$ denoting the matrix elements:  
	\begin{equation}
	C_{JL}^{(0)}=\frac{\sqrt{J(J+1)}L}{2\mu R^{2}},
	\end{equation}
	\begin{equation}
	C_{JL}^{(1)}=\frac{\sqrt{J(J+1)-2}L}{2\mu R^{2}},
	\end{equation}
	\begin{equation}
	C_{JL}^{(2)}=\frac{\sqrt{J(J+1)-6}L}{2\mu R^{2}},
	\end{equation}
	\begin{equation}
	C_{JS}^{(0)}=\frac{\sqrt2\sqrt{J(J+1)}}{2\mu R^{2}},
	\end{equation}
	\begin{equation}
	C_{JS}^{(1)}=\frac{\sqrt2\sqrt{J(J+1)-2}}{2\mu R^{2}},
	\end{equation}
	\begin{equation}
	C_{JS}^{(2)}=\frac{\sqrt2\sqrt{J(J+1)-6}}{2\mu R^{2}},
	\end{equation}
	\begin{equation}
	C_{LS}=\frac{\sqrt{2}L}{2\mu R^{2}},
	\end{equation}
	where $L$ is defined by the matrix elements of the $L^{\pm}$ operators:
	\begin{equation}
	L=\langle\Phi^{S\Sigma\Lambda+1}_{\Omega+1}|L_{+}|\Phi^{S\Sigma\Lambda}_{\Omega}\rangle=\langle\Phi^{S\Sigma\Lambda}_{\Omega}|L_{-}|\Phi^{S\Sigma\Lambda+1}_{\Omega+1}\rangle,
	\end{equation}
	We combine all matrices from Eq.~\ref{4}-\ref{12} into a big 20x20 matrix, which includes spin-orbit, Coriolis, and Gaussian couplings. Therefor, the full Hamiltonian matrix is expressed as follows:\\
	\onecolumngrid
	\newpage
	$\mathbf{H}= \\
	\\
	\fontsize{5}{6}
	\left[\begin{matrix}
	H_{diag}^{^{3}\pi}-A_{sd}/2&-\sqrt{3/2} A_{sd}+C_{LS}&0&0&0&0&0&0&-C_{JL}^{(0)}&-C_{JS}^{(0)}\\
	-\sqrt{3/2}A_{sd}+C_{LS}&H_{diag}^{^{3}\Sigma^{+}}&0&G^{(2)}&0&0&0&0&0&-C_{JL}^{(0)}\\
	0&0&H_{diag}^{^{3}\pi}-A_{sp}/2&-\frac{A_{sp}}{\sqrt{2}}+C_{LS}&0&0&0&0&0&0\\
	0&G^{(2)}&-\frac{A_{sp}}{\sqrt{2}}+C_{LS}&H_{diag}^{^{3}\Sigma^{+}}&0&0&0&0&0&0\\
	0&0&0&0&H_{diag}^{^{3}\pi}-A_{sd}/2&-\sqrt{3/2} A_{sd}&0&0&-C_{JL}^{(0)}&-C_{JS}^{(0)}\\
	0&0&0&0&-\sqrt{3/2} A_{sd}&H_{diag}^{^{1}\Sigma^{+}}&0&G^{(1)}&0&0\\
	0&0&0&0&0&0&H_{diag}^{^{3}\pi}-A_{sp}/2&-\frac{A_{sp}}{\sqrt{2}}&0&0\\
	0&0&0&0&0&G^{(1)}&-\frac{A_{sp}}{\sqrt{2}}&H_{diag}^{^{1}\Sigma^{+}}&0&0\\
	-C_{JL}^{(0)}&0&0&0&-C_{JL}^{(0)}&0&0&0&H_{diag}^{^{3}\Delta}-A_{sd}&A_{sd}/\sqrt{2}+C_{LS}\\
	-C_{JS}^{(0)}&-C_{JL}^{(0)}&0&0&-C_{JS}^{(0)}&0&0&0&A_{sd}/\sqrt{2}+C_{LS}&H_{diag}^{^{3}\pi}\\
	0&0&0&0&0&-C_{JL}^{(0)}&0&0&A_{sd}/\sqrt{2}&A_{sd}/2\\
	-C_{JL}^{(0)}&-C_{JS}^{(0)}&0&0&-C_{JL}^{(0)}
	&0&0&0&0&\sqrt{3}A_{sd}/2+C_{LS}\\
	0&0&-C_{JS}^{(0)}&-C_{JL}^{(0)}&0&0&-C_{JS}^{(0)}&0&0&0\\
	0&0&0&0&0&0&0&C_{JL}^{(0)}&0&0\\
	0&0&-C_{JL}^{(0)}&-C_{JS}^{(0)}&0&0&-C_{JL}^{(0)}&0&0&0\\
	0&0&0&0&0&0&0&0&-C_{JS}^{(1)}&-C_{JL}^{(1)}\\
	0&0&0&0&0&0&0&0&0&0\\
	0&0&0&0&0&0&0&0&0&-C_{JS}^{(1)}\\
	0&0&0&0&0&0&0&0&0&0\\
	0&0&0&0&0&0&0&0&0&0\\
	\end{matrix}......
	\right.
	\\
	\left.
	\fontsize{5}{6}
	......\begin{matrix}
	0&-C_{JL}^{(0)}&0&0&0&0&0&0&0&0\\
	0&-C_{JS}^{(0)}&0&0&0&0&0&0&0&0\\
	0&0&-C_{JS}^{(0)}&0&-C_{JL}^{(0)}&0&0&0&0&0\\
	0&0&-C_{JL}^{(0)}&0&-C_{JS}^{(0)}&0&0&0&0&0\\
	0&-C_{JL}^{(0)}&0&0&0&0&0&0&0&0\\
	-C_{JL}^{(0)}&0&0&0&0&0&0&0&0&0\\
	0&0&-C_{JS}^{(0)}&0&-C_{JL}^{(0)}&0&0&0&0&0\\
	0&0&0&-C_{JL}^{(0)}&0&0&0&0&0&0\\
	A_{sd}/\sqrt{2}&0&0&0&0&C_{JS}^{(1)}&0&0&0&0\\
	A_{sd}/2&\sqrt{3}A_{sd}/2+C_{LS}&0&0&0&-C_{JL}^{(1)}&0&-C_{JS}^{(1)}&0&0\\
	H_{diag}^{^{1}\pi}&-\sqrt{3}A_{sd}/2&0&0&0&0&-C_{JL}^{(1)}&0&0&0\\
	-\sqrt{3}A_{sd}/2&H_{diag}^{^{3}\Sigma^{+}}&0&0&G^{(2)}&0&0&-C_{JL}^{(1)}&0&0\\
	0&0&H_{diag}^{^{3}\pi}&-A_{sp}/2&\frac{A_{sp}}{2}+C_{LS}&0&0&0&-C_{JL}^{(1)}&0\\
	0&0&-A_{sp}/2&H_{diag}^{^{1}\pi}&\frac{A_{sp}}{2}&0&0&0&0&0\\
	0&G^{(2)}&\frac{A_{sp}}{2}+C_{LS}&\frac{A_{sp}}{2}&H_{diag}^{^{3}\Sigma^{+}}&0&0&0&-C_{JL}^{(1)}&0\\
	0&0&0&0&0&H_{diag}^{^{3}\Delta}&A_{sd}&A_{sd}/\sqrt{2}+C_{LS}&0&-C_{JS}^{(2)}\\
	-C_{JL}^{(1)}&0&0&0&0&A_{sd}&H_{diag}^{^{1}\Delta}&-A_{sd}/\sqrt{2}&0&0\\
	0&-C_{JL}^{(1)}&0&0&0&A_{sd}/\sqrt{2}+C_{LS}&-A_{sd}/\sqrt{2}&H_{diag}^{^{3}\pi}+\frac{A_{sd}}{2}&0&-C_{JL}^{(2)}\\
	0&0&-C_{JS}^{(1)}&0&-C_{JL}^{(1)}&0&0&0&H_{diag}^{^{3}\pi}+A_{sp}/2&0\\
	0&0&0&0&0&-C_{JS}^{(2)}&0&-C_{JL}^{(2)}&0&H_{diag}^{^{3}\Delta}+A_{sd}
	\end{matrix}
	\right].$
\end{document}